# USING DISENTANGLED STATES AND ALGORITHMIC INFORMATION THEORY TO CONSTRUCT A NOT P PROBLEM

RUBENS VIANA RAMOS

rubens@deti.ufc.br

*Department of Teleinformatic Engineering – Federal University of Ceara - DETI/UFC*

*C.P. 6007 – Campus do Pici - 60755-640 Fortaleza-Ce Brazil*

In this work, are used Chaitin's number $\Omega$ and the fact that the general decomposition of an *N*-way disentangled state is an irreducible sentence whose number of coefficients grows in a non-polynomial way with *N*, to construct a problem that can never be solved in P.

## 1. Introduction

A crucial problem in computer science is how efficient a problem can be solved in a computer [1 and references therein]. Briefly, a problem is said to be efficiently solved if the time required by a computer to find the correct solution grows in a polynomial way with the size of the input data. This class of problem is named P. However, there are some problems whose solutions, up to the moment, require a computational time that grows in a non-polynomial way with the size of the input data. The simplest class of this kind of problem is the NP class. The hardest of a NP problem is not to find a possible solution, but the number of possible solutions to be tested. This number grows exponentially with the growth of the size of the input data. However, if one has a probabilistic computer that always guess the correct answer and this one can be checked in a polynomial time, then the problem can be solved in polynomial time. The most popular NP problem is the travel salesman problem (TSP).

On the other hand, there are problems that are harder to solve than NP problems. Beyond the huge amount of solutions to be tested, the time required to check if a guessed solution is the correct one, also grows in a non polynomial way. In this direction, this work presents a problem that belongs to this class and it is shown that it can never be transformed to a P problem. In order to do this, it is made use of Chaitin's number $\Omega$ [2-4]. It is well known from the algorithmic information theory [5-7] that $\Omega$, a infinite binary sequence, is irreducible. This means that the smallest program able to provide the firsts *L* bits of $\Omega$ has length of about *L* bits. Here, it is proposed a problem whose solution requires the knowledge of the firsts *L* bits of $\Omega$, with *L* growing in a non-polynomial way with the size of the input data. This means that the length of the program able to solve the proposed problem will also grow in a non-polynomial way, as well the time required by the computer to run the program and find the answer. The proposed problem makes use of the general decomposition of an *N*-way disentangled quantum state. This work is outlined as follows: in Section 2 the general decomposition of an *N*-way disentangled quantum state is reviewed; in Section 3 the problem is proposed and, at last, the conclusions are presented in Section 4.

## 2. *N*-way disentangled States

An *N*-way disentangled state is an *N*-partite (of qubit) quantum state that has not simultaneously non-local correlation between all *N* individual subsystems. The general decomposition of an *N*-way disentangled state consists of a mixture of pure *N*-way disentangled states. In other words, a mixture of pure *N*-way disentangled states can not produce an *N*-way entangled state. In order to see this point, it is useful to use the entanglement of formation [8-11]. The *N*-way entanglement of a *N*-partite quantum state $\Gamma_{12..N}$ is given by

$$E_F\left(\Gamma_{12...N}\right) = \min \sum_i p_i E\left(\gamma^i_{12...N}\right) \bigg| \Gamma_{12...N} = \sum_i p_i \gamma^i_{12...N}, \qquad (1)$$

where $\gamma^i_{12...N}$ is a *N*-partite pure quantum state and *E* is an adequate entanglement measure for pure *N*-partite quantum states. If $\gamma^i_{12...N}$ are all disentangled, then $E\left(\gamma^i_{12...N}\right) = 0 \ \forall i$ and, hence, $E_F\left(\Gamma_{12...N}\right) = 0$. For example, a bipartite state $\Gamma_{12}$ has 2-way entanglement if

$$\Gamma_{12} \neq \sum_i p_i \left(\rho^i_1 \otimes \rho^i_2\right) \qquad (2)$$

$$\sum_i p_i = 1 \qquad (3)$$

where $\rho_1$ and $\rho_2$ are pure single-qubit states. On the other hand, a tripartite state $\Gamma_{123}$ has 3-way entanglement if

$$\Gamma_{123} \neq \sum_i p_i \left(\rho^i_1 \otimes \rho^i_2 \otimes \rho^i_3\right) + \sum_j r_j \left(\rho^j_1 \otimes \Phi^j_{23}\right) + \sum_l q_l \left(\rho^l_2 \otimes \Phi^l_{13}\right) + \sum_k t_k \left(\rho^k_3 \otimes \Phi^k_{12}\right) \qquad (4)$$

$$\sum_i p_i + \sum_j r_j + \sum_l q_l + \sum_k t_k = 1 \qquad (5)$$

where $\Phi_{12}$, $\Phi_{23}$ and $\Phi_{13}$ are entangled pure bipartite states. When the number of qubits *N* increases, the number of terms (summatories) of the general *N*-way disentangled state grows in a non-polynomial way. In fact it is easy to show that for *N*>4 the number of terms is larger than $2^N$. Each term has a finite number of coefficients. For example, for a bipartite state, equation (2), it is known that the summatory has 16 coefficients [10]. Hence, the total number of coefficients of a general decomposition of an *N*-way disentangled state grows really fast when *N* grows. Since each pure quantum state of the general *n*-way disentangled state decomposition represents physically a different kind of quantum state, in the sense that the positions of the entanglement are different, the sentence of the general decomposition of an *N*-way disentangled state cannot be reduced to a smaller one containing a lower number of terms. For a tripartite state, for example, it is obvious that

$$\rho_1 \otimes \rho_2 \otimes \rho_3 \neq \rho_1 \otimes \Phi_{23} \neq \rho_2 \otimes \Phi_{13} \neq \rho_3 \otimes \Phi_{12} \quad \forall \rho_1, \rho_2, \rho_3, \Phi_{12}, \Phi_{13}, \Phi_{23} \tag{6}$$

and equation (4) can never has less than 4 terms. Thus, for instance, one can not construct a 3-way pure disentangled state having 2-way entanglement between subsystems 2 and 3, using pure completely disentangled tripartite states or pure tripartite states having 2-way entanglement between subsystems 1 and 3. Hence, the general decomposition of an *N*-way disentangled state is an irreducible sentence, in the same sense that Chaitin's number $\Omega$ is an irreducible binary number.

Since the number of terms of the general decomposition of an *N*-qubit disentangled state grows in a non-polynomial way, several problems where it is necessary and the number of qubits increases are problems that are not in P. In the next section it is presented an artificial problem based on the general decomposition of an *N*-way disentangled state that can not be reduce to P.

## 3. A not P problem using the general decomposition of an *N*-way disentangled States and $\Omega$

Let us suppose the following problem: Find any *N*-way disentangled state $\Phi_\Omega$ whose coefficients of the general decomposition $b_i$ ($\sum_i b_i = 1$) are obtained from the bits of Chaitin's number $\Omega$ as explained below:

1. *S* is the total number of coefficients (that grows in a non-polynomial way with *N*).
2. $\Omega_{ST} = c_1 c_2 c_3 c_4 \ldots c_{ST}$ are the firsts *ST* bits of $\Omega$, where $T = \lceil \log_2(S) \rceil$ (the first integer larger than $\log_2(S)$).
3. $b_i = B_i / K_b$ where $K_b = \sum_i B_i$ and $B_i$ is the decimal value of the *i*-th sequence of *T* bits of $\Omega$. For example, the value of $B_1$ is bin2dec($c_1 c_2 \ldots c_T$), $B_2$ is bin2dec($c_{T+1} c_{T+2} \ldots c_{2T}$) and so on. The function 'bin2dec' calculates the decimal value of the bit sequence in its argument.

Given the number of qubits *N*, the value that each $B_i$ can assume belongs to the interval [0-(*S*-1)]. The sequence of integer numbers $B_i$, whose concatenated binary representation resembles $\Omega_{ST}$ is the solution. Obviously the solution can be found using the brute force method by testing all possible sequences of $S$ integer numbers where each integer number can assume $S$ different values. On the other hand, a very lucky probabilistic computer could try to solve problem in polynomial time guessing correctly the sequence of integer numbers $B_i$. However, the proposed problem is a not a NP problem because once the solution was guessed, it would take a non-polynomial time to check it. In order to show that the proposed problem can not be solved in a polynomial time by deterministic or probabilistic computers, one must consider that, if that was possible, it would mean that it is always possible to find $\Omega_{ST}$ (the firsts $ST$ bits of $\Omega$, with $T=\log_2(S)$ and $S$ a function of $N$) in a polynomial time. However, according to the algorithmic information theory, the smallest program able to provide $\Omega_{ST}$ has length almost equal to the length of $\Omega_{ST}$. Since $S$ grows in a non-polynomial way with $N$, the length of $\Omega_{ST}$ and of the smallest program able to provide it, also grow in a non-polynomial way. Hence, the time required to generate $\Omega_{ST}$ grows in a non-polynomial way with $N$. In other words, the length of the smallest program able to find the solution of the proposed problem and the time to run it grow, unavoidable, in a non-polynomial way with $N$. Now supposes that an arbitrarily large amount of bits of $\Omega$ is stored in an arbitrarily large amount of memory and all that the computer has to do in order to obtain $\Omega_{ST}$ is, instead of calculate, to read the necessaries bits of $\Omega$ in the memory. Since the length of $\Omega_{ST}$ grows in a non-polynomial way, the time required to read $\Omega_{ST}$ will also grow in a non-polynomial way and, hence, the smallest program to solve the proposed problem requires a time that grows in a non-polynomial way with $N$.

## 4. Conclusions

It was shown that the problem of constructing any $N$-way disentangled state, whose coefficients of the general decomposition are obtained from Chaitin's number as explained in the text, is a not P problem and the program able to solve it necessarily requires a time that grows in a non-polynomial way with the number of qubits $N$.

## Acknowledgements

The author thanks Paulo Benício Melo for the useful comments in computational complexity.